\documentclass{aa}
\usepackage{graphicx}
\usepackage{txfonts}
\usepackage{amsmath}
\usepackage{amsfonts} 
\usepackage{lscape}

\def\teff{${\rm T_{\rm eff}}$}
\def\gr{$\log {\rm g}$}
\def\vt{${\rm v_{\rm turb}}$}
\def\alfa{[$\alpha$/Fe]}
\def\tros{$\log{\tau_{Ross}}$}

\begin{document}

\title{KOALA, a new ATLAS9 database} 
\subtitle{I. Model atmospheres, opacities, fluxes, \\ bolometric corrections, magnitudes and colours}

\author{A. Mucciarelli\inst{1,2}, P. Bonifacio\inst{3,4}, C. Lardo\inst{1,2}}

\institute{
Dipartimento  di  Fisica  e  Astronomia  “Augusto  Righi”,  Alma  Mater Studiorum, Universit\`a  di Bologna, Via Gobetti 93/2, I-40129 Bologna, Italy
\and
INAF - Osservatorio di Astrofisica e Scienza dello Spazio di Bologna, Via Gobetti 93/3, I-40129 Bologna, Italy;
\and
LIRA, Observatoire de Paris, Universit{\'e} PSL, Sorbonne Universit{\'e}, Universit{\'e} Paris Cité, CY Cergy Paris Universit{\'e}, CNRS,92190 Meudon, France
\and
INAF - Osservatorio Astronomico di Trieste, Via G.B.Tiepolo 11, Trieste, I-34143, Italy
}

\authorrunning{Mucciarelli et al.}
\titlerunning{ATLAS9 model atmospheres}

\abstract{
We present the KOALA database, a new set of LTE, line-blanketed model atmospheres calculated with the code ATLAS9, together 
with the corresponding Opacity Distribution Functions and emergent fluxes. The latter were used also to calculated 
G-band bolometric corrections and theoretical magnitudes and colours for several photometric systems, i.e. UBVRI, 2MASS, 
Hypparcos-Tycho, SDSS, Galex, Euclid and Gaia DR3. 
With respect to the previous grids of ATLAS9 model atmospheres, we adopted the solar mixture by Caffau/Lodders and 
we extend the sampling in metallicity (from --5.0 to --2.5 dex with step of 0.5 dex, and 
from --2.5 dex to +0.5 dex with step of 0.25 dex) and in \alfa\ 
(from --0.4 to +0.4 dex with a step of 0.2 dex). 
Also, we provide a finer sampling in \teff\ for \teff\ lower than 7000 K. 
This finer grid allows for more accurate interpolation of colours and in many cases it makes
not necessary to compute a new model atmosphere, since one of the grid can be used directly.
A total of 51663 model atmospheres and emergent fluxes have been computed. 
Finally, we discuss the impact of [M/H] and \alfa\ on the thermal and pressure structures of the 
model atmospheres and on theoretical colours.}
 
\keywords{stars: atmospheres }
               
\maketitle
%
%________________________________________________________________

\section{Introduction}

A model atmosphere describes how thermodynamic quantities (i.e. temperature, gas pressure, electron number density) 
vary as we move through the photosphere, from outer, optically thin layers to the deeper, optically 
thick regions where continuum photons are formed. 
The run of these quantities with the depth can be expressed as a function of the optical depth (in plane-parallel 
geometry) or of the stellar radius (in spherical geometry).  
These models are the backbone of the spectral synthesis and of the modelling 
of the spectral energy distribution of the stars, that we observe at different resolution 
through photometric filters or spectra. 
The relevance of the model atmospheres in astrophysics is multiple, being a fundamental 
element in the chemical abundance analysis, in the transformation of stellar tracks/isochrones 
into observational planes
and in the modelling of the spectral energy distribution of the galaxies.

ATLAS9 \citep{kurucz70,kurucz74,kurucz2005} and ATLAS12 \citep{kurucz2005,castelli05at12} 
are the only publicly available codes to calculate model atmospheres, together with TLUSTY 
\citep{hubeny88,hubeny95,hubeny03} and TMAP \citep{werner99}, while other codes are not 
publicly distributed despite the wide use of their grids of model atmospheres, for instance 
MARCS \citep{gust2008} and PHOENIX \citep{hauschildt97,hauschildt99}.
With respect to other codes that are suitable for specific regions of the parameter spaces, 
i.e. TLUSTY for OB stars \citep[but usable also for cooler stars, see e.g.][]{hubeny21}, MARCS for FGKM stars, PHOENIX for AFGKM stars, 
ATLAS9/ATLAS12 allow us to calculate reliable model atmospheres over a large range of parameters.
These models assume one-dimensional, plane-parallel geometry, 
local thermodynamic and radiative equilibrium, and the radiation/convection as the 
only energy transport mechanisms. 
The main difference between ATLAS9 and ATLAS12 is in the treatment of the line opacity. 
ATLAS9 handles the complex and time-consuming step of the calculation of the 
line opacity coefficient using the opacity distribution function (ODF) method \citep[see][and references therein]{kurucz79}. 
In this approach the line opacity for a given chemical mixture and microturbulent velocity (\vt)
is calculated as a function of temperature and gas pressure in a number of wavelength intervals, 
namely 328 for the so-called big ODF (used in the model atmosphere calculation) and 
1212 for the so-called little ODF (used in the emergent flux calculation).

The main advantage of the ODF method is that, after a relatively time-consuming effort 
to calculate big and little ODFs, any model atmosphere or emergent flux with the chemical mixture 
and \vt\ of the corresponding ODF can be calculated in a very short time, adopting 
the appropriate \teff\ and \gr . This approach is therefore suitable to calculate 
a large number of models and emergent fluxes for stars with similar chemical composition.
For stars with peculiar chemical compositions (or with specific elemental abundances that significantly 
impact on the total opacity) the opacity sampling method \citep[][and implemented in ATLAS12]{peytremann1974} 
is recommended allowing us the proper calculation of the opacity coefficients for the parameters 
of the model.

The availability of extensive grids of atmospheric models (as well as ODFs and fluxes) that
adequately sample the parameter space (\teff , \gr, but also the chemical mixture) 
is fundamental for various aspects of astrophysics. 
The most used grids of ATLAS9 model atmospheres have been provided 
by \citet[][hereafter CK03]{ck03}\footnote{https://wwwuser.oats.inaf.it/fiorella.castelli/grids.html}. 
This grid assumes as reference solar mixture 
that by \citet{gs98} and provides metallicities [M/H] between --4.0 and +0.5 dex, with a typical step 
of 0.5 dex, and two values of \alfa\, +0.0 and +0.4 dex. However, the grid is not sampled in a regular way, 
lacking with some combination of [M/H] and \alfa. 

Here we present the first paper of the KOALA project, a new database of 
opacities, model atmospheres and emergent fluxes, calculated with the ATLAS9 
software. In this paper we describe the new grid that substitutes the previous 
ones by CK03 and following the same physical assumptions, in particular for the 
adopted opacities. In the next papers we will discuss future grids testing 
the impact of peculiar chemical composition and of some physical assumptions.

\section{The new ATLAS9 grid}

We present new, wider and complete grid of ODFs, model atmospheres and emergent fluxes, 
calculated with the {\tt gfortran} 
version of the codes DFSYNTHE, KAPPA9 \citep{castelli05df} and ATLAS9 \citep{kurucz2005}, 
all of them available in the F. Castelli 
website\footnote{https://wwwuser.oats.inaf.it/castelli/sources/atlas9codes.html}.
With respect to the grid by CK03, we adopted a different solar chemical composition, 
we increase the parameter sampling in [M/H] and \alfa\ and we increase 
the sampling in \teff\ for the cool models (see details in Section~\ref{novel}). 
All the products of this new release (big and little ODFs, Rosseland opacity tables, model atmospheres, 
emergent fluxes, G-band bolometric corrections and theoretical magnitudes) are available in a dedicated web-site\footnote{https://sites.google.com/view/koala-database/}, 
where we will provide future, new grids of models calculated with specific chemical mixtures or 
testing different physical assumptions. 
The dataset presented here has been already used to calculate the 
colour-temperature transformations used to transform into observative planes 
the theoretical isochrones and tracks of the BaSTI-IAC database\footnote{http://basti-iac.oa-abruzzo.inaf.it/} \citep{hidalgo18,pietrinferni21,pietrinferni24}.

\subsection{Main features and novelties}
\label{novel}

In the following we summarise the characteristics of this new grid and the 
main improvements with respect to the previous one by CK03.
\begin{itemize}
\item 
The reference solar chemical mixture is composed by the solar abundances by \citet{caffau11} for Li, C, N, O, P, S, K, Fe, 
Eu, Hf, Os and Th, and those by \citet{lodders10} for the other elements.
This new solar chemical mixture substitutes that by \citet{gs98} adopted by CK03. 
For all the ODFs we assumed a helium mass fraction of Y=0.2476, following the {\sl Planck} mission results \citep{coc14}. 
We point out that the relatively small change in He abundance due to the Galactic chemical
evolution has minor effects on the structure of the model atmospheres. For a detailed description
on the effects of the He abundance on the spectra of latte-type stars we refer the reader to \citet{vitense79}. 
\item 
We considered different chemical mixtures in terms of $\alpha$-elements (O, Ne, Mg, Si, S, Ar, Ca and Ti), 
namely \alfa =--0.4, --0.2, +0.0, +0.2 and +0.4 dex. 
The grid by CK03 included only \alfa =+0.0 and +0.4 dex. 
\item 
For each value of \alfa , we consider 18 metallicities. 
The adopted metallicities range from [M/H]=--5.0 dex to --2.5 dex with a step of 0.5 dex, 
and from --2.5 dex and +0.5 dex with a step of 0.25 dex.
This significantly extends the metallicity range and sampling of the grid by CK03, the latter 
adopting a sampling of +0.5 dex and with some lacking metallicities for \alfa=+0.4 dex.
Therefore, for each \alfa\ we provide a regular grid of models in terms of [M/H].
\item 
Another novelty of the new grid with respect to the grid by CK03 is the adoption 
of a step of 125 K in the range 3750-6000 K (for \gr$\le$3.0) and 
in the range 3750-7000 K (for \gr$>$3.0), finer by a factor of 2 with respect to the previous grid.
This finer grid allows for more accurate interpolation of colours and in many cases it makes
not necessary to compute a new model, since one of the grid can be used directly.
We add as boundary of the grid models with \teff=3750 K even if problems in the ATLAS9 integrated 
colours for \teff\ lower than 4000 K has been pointed out \citep[see e.g.][]{plez11}. 
For hotter \teff\ we maintain the same step by CK03, 250 K for stars up to 
\teff\ =12000 K, 500 K up to \teff\ =20000 K and 1000 K up to \teff\ =50000 K  . 
Also for \gr\ we adopted the same step (0.5 dex) by CK03. 
Fig.~\ref{grid} shows the sampling of the entire grid in the \teff\--\gr\ plane, compared with the 
extension of the MARCS \citep{gust2008}, PHOENIX \citep{husser13} and TLUSTY \citep{lanz03,lanz07} 
model atmospheres.

\begin{figure}[ht!]
\centering
\includegraphics[scale=0.35]{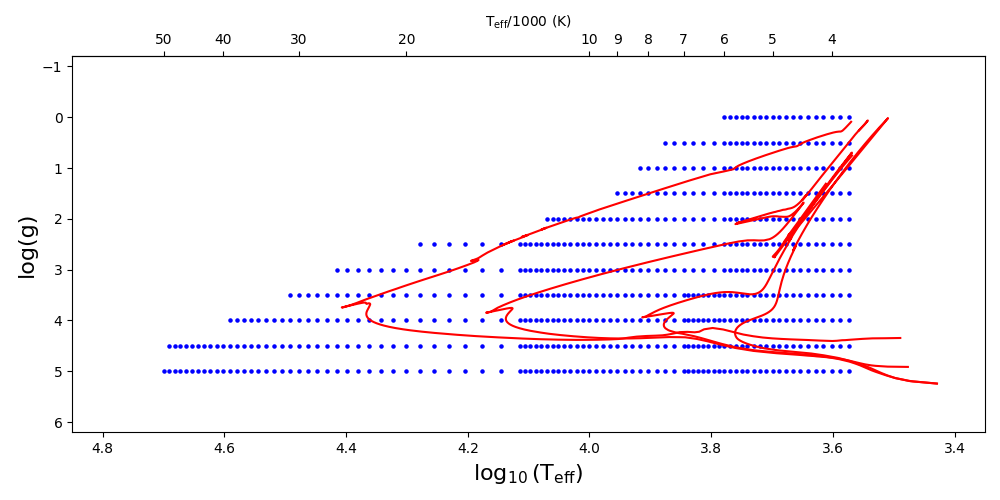}
\includegraphics[scale=0.35]{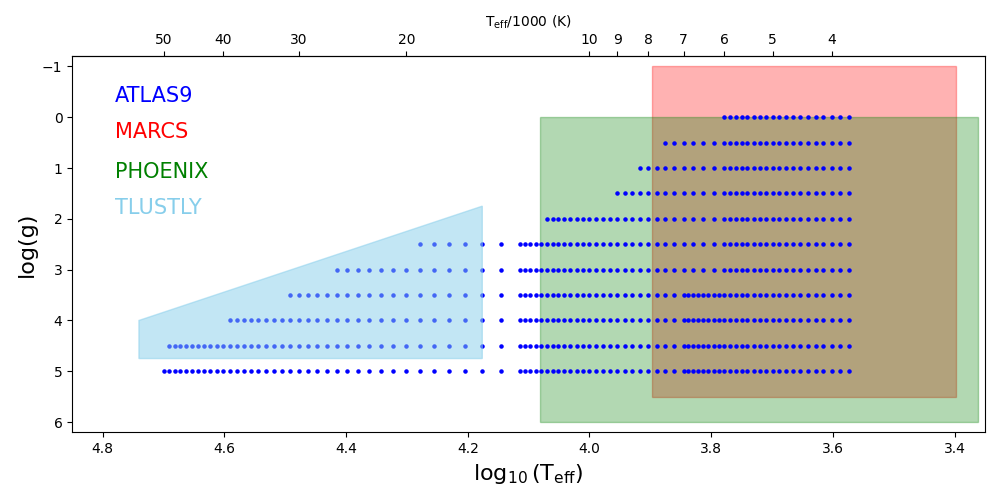}
\caption{Distribution of the ATLAS9 model atmospheres of the new grid (blue circles) in the $\rm log_{10}$(\teff) --\gr\ plane, 
superimposed (upper panel) to four theoretical BaSTI-IAC isochrones with 15 Myr, 100 Myr, 1 Gyr and 13 Gyr 
and solar-scaled metallicity \citep{hidalgo18}, and compared (lower panel) to the distribution of 
MARCS \citep{gust2008}, PHOENIX \citep{husser13} and TLUSTY \citep{lanz03,lanz07} model atmospheres.}
\label{grid}
\end{figure}

\end{itemize}

\subsection{Calculation of ODFs, model atmospheres and emergent fluxes}
\label{calculation}

The ODFs were computed with DFSYNTHE and KAPPA9 codes \citep{castelli05df}, adopting the same atomic and molecular line lists used by CK03 and described in \citet{kurucz11}. 
Briefly, in the computation of ODFs we included all the atomic lines in the last release of the 
R. L. Kurucz dataset, together with the linelists for several diatomic molecules, i.e. 
$\rm H_2$, CH, NH, OH, MgH, SiH, $\rm C_2$, CN, CO, SiO, and TiO 
\citep[for the latter we adopted the linelists by][]{schwenke98}.
The only difference with respect to the opacities used by CK3 is related to $\rm H_{2}O$.
For the latter we adopted the last version of the \citet{ps97} linelist 
as provided by R. L. Kurucz\footnote{http://kurucz.harvard.edu/molecules/h2o/}, fixing a bug present in the previous release of the same linelist.  In the future releases of KOALA, we plan to improve new 
opacity sources not included in the last release of Kurucz's database.
For each chemical composition, specified by the metallicity [M/H] (see Appendix~\ref{metkur}) and the $\alpha$-element 
abundance ratio \alfa, we calculated big and little ODFs adopting five values of \vt , namely 0, 1, 2, 4 and 8 km/s . 

Model atmospheres with \vt=2 km/s were calculated with ATLAS9 \citep{kurucz2005}, adopting 
the big ODFs.  
ATLAS9 enforces Saha–Boltzmann ionization/excitation plus molecular dissociation equilibrium at each depth, together with elemental abundance conservation and charge neutrality. The code  solves a coupled set of equilibrium equations for all species (atoms/ions and selected molecules), returning number densities and the electron pressure for a given 
temperature, pressure and composition.
For additional details about the main physical assumptions and the calculation scheme adopted by ATLAS9 
we refer the reader to the vast literature available on this code 
\citep{kurucz70, kurucz74, kurucz79, castelli88}. 
Since a model atmosphere is calculated through an iterative numerical process 
starting from a guess solution, these models were obtained by choosing as the guess model 
the CK03 grid model closest in terms of \teff\ , \gr\ , and [M/H] 
to the solution to be computed. The choice of a different guess model, as long as it is close to
the desired solution, has a negligible impact on the computed model, whereas using guess models 
that are far from the required solution can slow down the calculation or compromise the convergence 
of the final model.

All of the models have 72 plane-parallel layers ranging from the logarithm of the Rosseland 
optical depth \tros  = -6.875 to +2.00, 
in steps of $\delta$\tros\ =0.125. In the treatment of the convective flux, 
the classical mixing length theory \citep{vitense53,vitense58} is adopted, 
with the pressure scale height equal to $\alpha$ times the characteristic length 
of a rising convective cell. Here we adopted 
a mixing length parameter $\alpha$ equal to 1.25, as done in previous grids of ATLAS9 
models \citep[CK03,][]{kirby2011,meszaros2012}. We stress as the adoption of a different value impact only 
on the deepest layers of the coolest models, where continuum and very weak lines form.
The approximate overshooting \citep{kurucz92} is switched off because, even if it is able to better reproduce 
some spectral features in the solar spectrum, for most of the stars it does not 
provide a good description of several features \citep{castelli97} or introduces unphysical features in the 
thermal structure of the most metal-poor models \citep{bonifacio09}. 
For all the layers, local thermodynamical equilibrium is assumed 
\citep[but this assumption can becomes progressively less realistic 
for temperatures hotter than $\sim$15000 K, see e.g.][]{lanz07}.

For any model, a given layer is considered converged if its flux and flux derivative 
errors are less than 1\% and 10\% , respectively. 
The majority of the computed models satisfies these criteria in all the layers. 
In some models (especially those with low \teff\ and \gr) some layers are not converged, 
usually the most external ones. 
Sometimes, small changes 
in the guess model atmosphere (for instance of the order of some tens of K) can be enough to solve 
the problems of convergence of a layer. However, possible not converged outermost layers do not 
affect significantly the use of the model atmosphere, because in those layers the core of strong lines 
formed, usually not adopted for chemical analyses because of some effects (i.e., non-LTE, chromospheric activity) 
are not accounted for in ATLAS models. 
Additionally, for some low \teff ($\lesssim$4600), high \gr\ ($\ge$2.0-2.5) 
and low [M/H] models ([M/H]$<$--2.5 dex, i.e. models corresponding 
to metal-poor, low Main Sequence stars, as K and M dwarfs) we found problems of convergence 
also in the central layers, where a significant fraction of the flux is transported by convection 
(in particular an unphysical discontinuity in temperature is present close to these layers).
The use of different guess model atmospheres or small changes in the requested parameters were not 
able to fix this issue and we decide to remove these models from the grid. 

The number of model atmospheres of the final grid is summarised in Table~\ref{nmodel}, 
according to the adopted [M/H] and \alfa . A total of 51663 model atmospheres is 
provided.
Finally, for each model of the grid described above the corresponding emergent flux 
(i.e. a low-resolution spectrum covering the entire spectral range) has been calculated 
as $\rm H_{\nu}$, energy for frequency unit.

\tiny
\begin{table*}
\caption{Number of computed model atmospheres and fluxes according to 
the metallicity and \alfa .}            
\label{nmodel}     
\centering                          %
\begin{tabular}{c   c c c  c c}       
\hline\hline                 % inserts double horizontal lines
[Fe/H] &   \alfa=--0.4 & \alfa=--0.2  & \alfa=+0.0 & \alfa=+0.2  & \alfa=+0.4   \\
\hline
  (dex) &  &  &  &  &  \\
\hline                        % inserts single horizontal 
 +0.50        &     579         &      579          &   579    &     579        &    579          \\	  
 +0.25        &     579         &      579          &   579    &     579        &    579          \\	  
 +0.00        &     579             &      579          &   579    &     579        &    579          \\	  
--0.25        &     579              &      579          &   579    &     579        &    579          \\	  
--0.50        &     579             &      579          &   579    &     579        &    579           \\	  
--0.75        &     579               &      579          &   579    &     579        &    579             \\	  
--1.00        &     579              &      579          &   579    &     579        &    579              \\	  
--1.25        &     579              &      579          &   579    &     579        &    579             \\	  
--1.50        &     579              &      579          &   579    &     579        &    579              \\	  
--1.75        &     579              &      579          &   579    &     579         &    579              \\	  
--2.00        &     579              &      579          &   579    &     579        &    579              \\	  
--2.25        &     579              &      579          &   579    &     579        &    579              \\	  
--2.50        &     578             &      579          &   579    &     579         &    579              \\	  
--3.00        &     575          &      573          &   576    &     565        &    568           \\	  
--3.50        &     565         &      567          &   567    &     564         &    568           \\	  
--4.00        &     559         &      559          &   562    &     561           &    561          \\	  
--4.50        &     555         &      559          &   555    &     553        &    556           \\	  
--5.00        &     554         &      551          &   550    &     551        &    555          \\
\hline
 TOTAL        &   10334           &   10336            &  10337   &    10321         &  10335   \\

\hline                                   %inserts single line
\end{tabular}
\end{table*}

\normalsize
\section{Impact of the adopted solar chemical mixture}
\label{solar}

\begin{figure*}[ht!]
\centering
\includegraphics[width=\hsize]{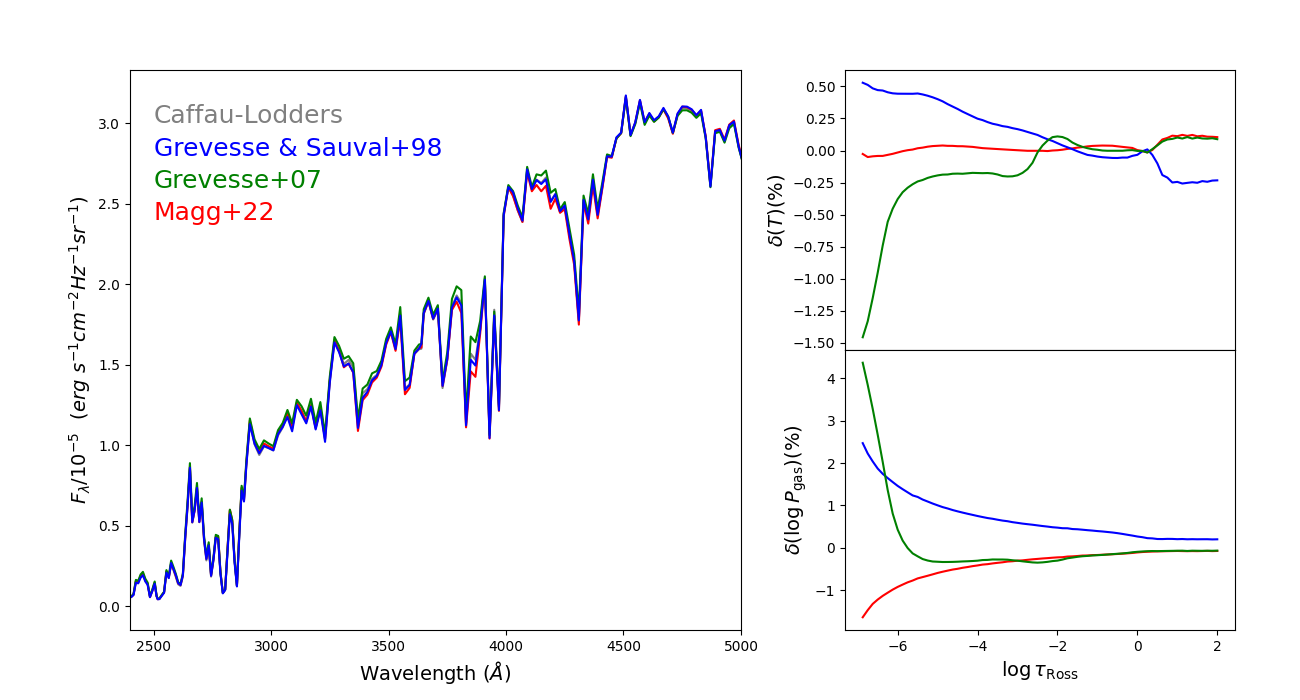}
\caption{Main panel: comparison among ATLAS9 fluxes for the Sun calculated with suitable 
ODFs adopting different solar chemical mixtures: Caffau-Lodders (this work, grey line), 
\citet[][blue line]{gs98}, \citet[][green line]{grevesse07}, \citet[][red line]{magg22}.
Left panels: percentage difference in temperature (upper panel) and logarithm of the gas pressure 
(lower panel) as a function of the Rosseland optical depth of the model atmospheres with respect to that 
computed with Caffau-Lodders chemical mixture (same colour-code of the main panel).
}
\label{sun}
\end{figure*}

The comparison between different families of model atmospheres is not trivial because of 
the large number of assumptions in the recipes adopted by different codes. 
In order to evaluate the true impact of the different solar chemical mixtures on the model atmospheres and fluxes, 
we calculated additional ODFs, assuming the solar mixtures by \citet{gs98}, used by CK03,  
\citet{grevesse07}, used in the MARCS models grid by \citet{gust2008} and the recent one by \citet{magg22}. 
The most significant differences among these four mixtures are for C, N and O.
In Fig.~\ref{sun} we compare the emergent fluxes and the thermal and pressure structures of model atmospheres 
for the Sun adopting these chemical mixtures. Small differences are appreciable in the outermost regions, 
especially for the model adopting the mixture by \citet{grevesse07}, characterized by the lowest 
C, N, O abundances. 
This exercise suggests that the adoption of the solar chemical mixture does not significantly impact 
on the structure of the model atmospheres. On the other hand, the adopted solar abundances 
have a significant impact on the derived fluxes in the UV/blue spectral regions dominated by 
strong CH, NH and CN molecular bands. 
Checks performed for other sets of metallicity and stellar parameters 
lead to similar conclusions.

\normalsize
\section{Impact of the metallicity on the model atmospheres}
\label{metallicity}

\begin{figure*}[ht!]
\centering
\includegraphics[scale=0.55]{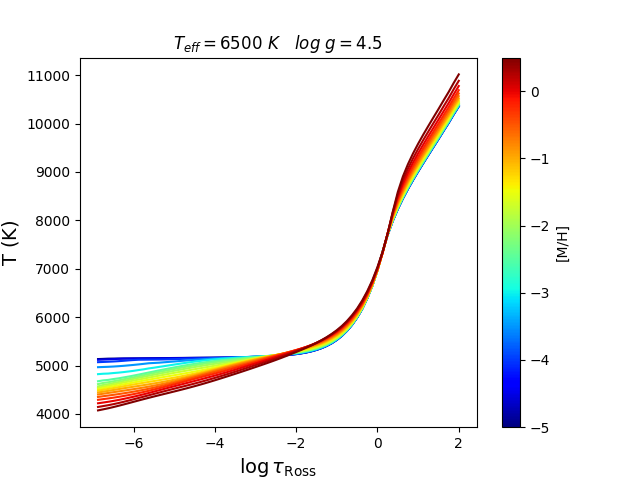}
\includegraphics[scale=0.55]{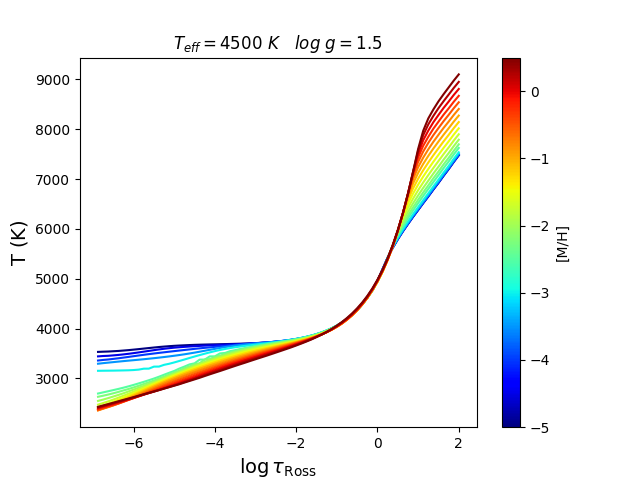}\\
\includegraphics[scale=0.55]{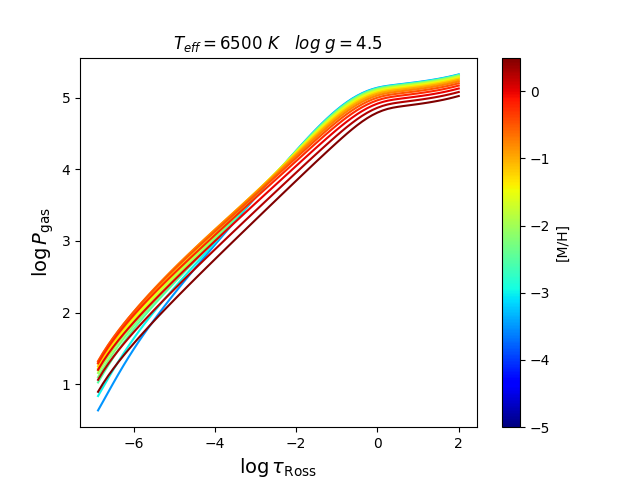}
\includegraphics[scale=0.55]{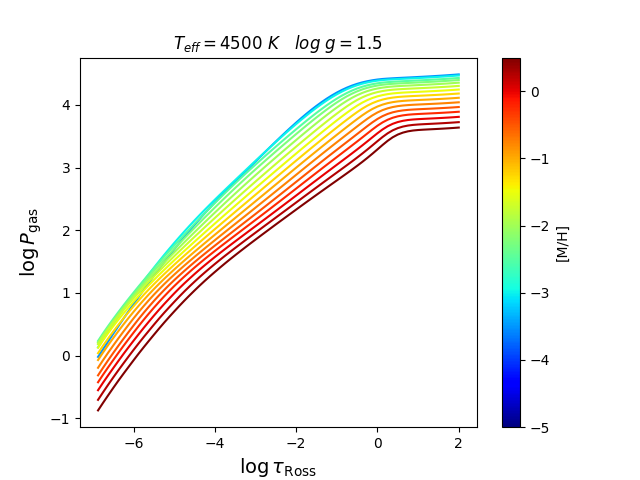}\\
\includegraphics[scale=0.55]{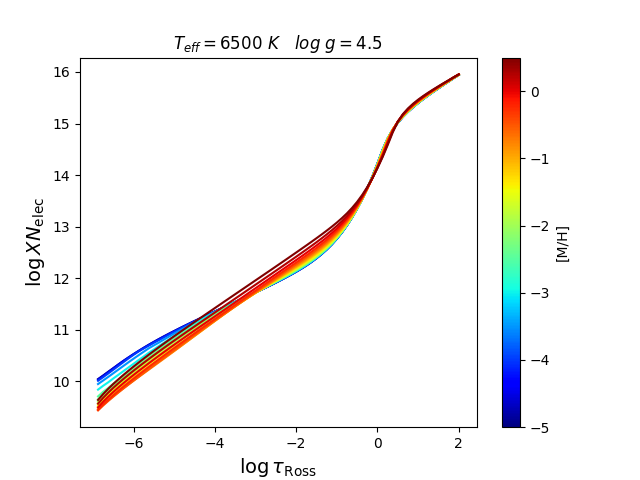}
\includegraphics[scale=0.55]{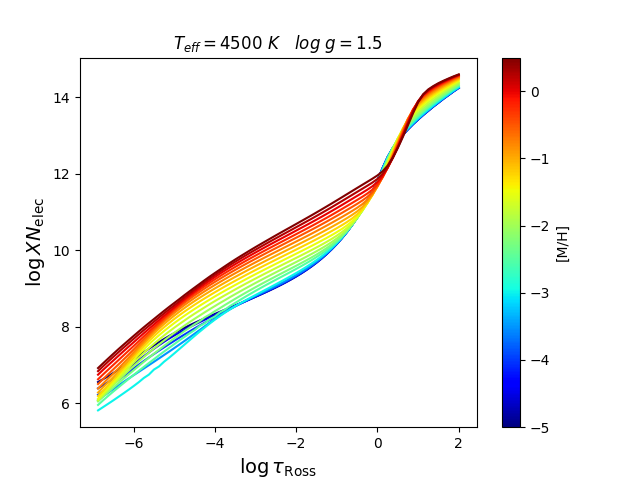}
\caption{Run of temperature (upper panels), logarithm of the gas pressure (middle panels) and 
logarithm of the electron number density (lower panels) as a function of \tros\ for the model atmospheres 
of a dwarf (\teff=6500 K and \gr=4.5, left panels) and a giant (\teff=4500 K and \gr=1.5, right panels) star, 
colour-coded according to the metallicity [M/H]. We adopt \alfa=+0.0 dex for all the models. }
\label{mod_teff}
\end{figure*}

Fig.~\ref{mod_teff} shows how model atmospheres for a giant (\teff=4500 K and \gr=1.5) 
and for a dwarf (\teff=6500 K and \gr=4.5) star change by changing [M/H]. In particular, we show the run 
of temperature, gas pressure ($\rm log(P_{gas})$) and electron number density ($\rm log(XN_{elec})$) as a function of \tros . 
It is immediately clear that models with [M/H]$\gtrsim$--2.5/--2.0 dex become 
increasingly distinct from each other at a fixed optical depth, while more metal-poor models are often 
indistinguishable or show only small differences. This justifies our choice of a finer sampling at higher 
metallicities, namely +0.25 instead of +0.5 adopted by CK03. 
As exercise, we discuss in Appendix~\ref{zero} the properties of a zero-metallicity ODF, representative 
of an ideal Population III star.

The thermal structure of the model is sensitive to [M/H] in particular in deeper (\tros$\gtrsim$0.5) 
and outer (\tros$\lesssim$-3) layers (with the differences more pronounced in the giant model).
Fixing \teff\ (and therefore the total flux, according to the Stefan-Boltzmann relation), 
a higher metallicity leads to a larger absorption from metallic lines and 
a higher opacity. This reduces the capability of the star to radiate energy and the temperature of the innermost layers 
must increase to make the total flux constant. On the other hand, the temperature of the outermost layers 
(where the atmosphere becomes optically thin) decreases because it is easier for the model to radiate 
\citep[see e.g.][]{kurucz79}.

Also the gas pressure and the electron number density are significantly affected by [M/H] 
but in different and opposite ways. 
For [M/H]$\gtrsim$--3.0/--2.5 dex, $\rm log(P_{gas})$ increases by decreasing [M/H] 
at each \tros , while $\rm log(XN_{elec})$ decreases.  
This occurs because, as metallicity decreases, the fraction of free electrons (provided by metals) 
also decreases leading to a lower electron number density (lower panels of Fig.~\ref{mod_teff}) 
and therefore a lower number of $H^{-}$ atoms. 
This corresponds to a lower line opacity coefficient that increases the 
gas pressure according to the hydrostatic equilibrium (middle panels of Fig.~\ref{mod_teff}) 
even if the electron pressure decreases. 

\begin{figure*}[ht!]
\centering
\includegraphics[scale=0.55]{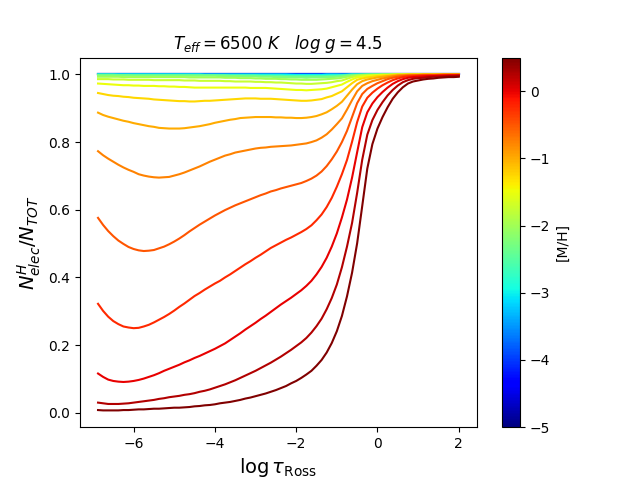}
\includegraphics[scale=0.55]{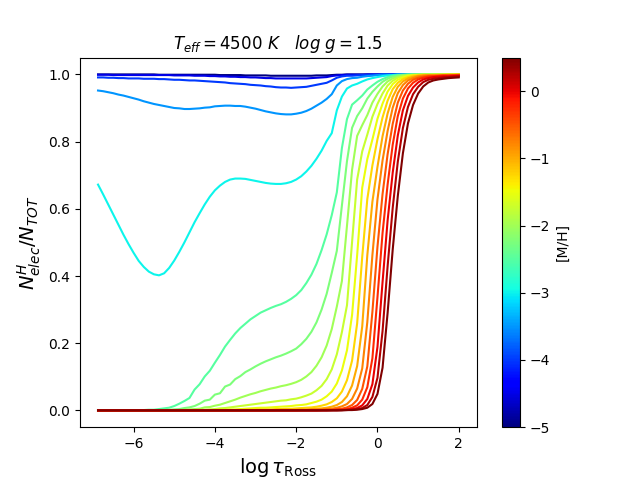}\\
\includegraphics[scale=0.55]{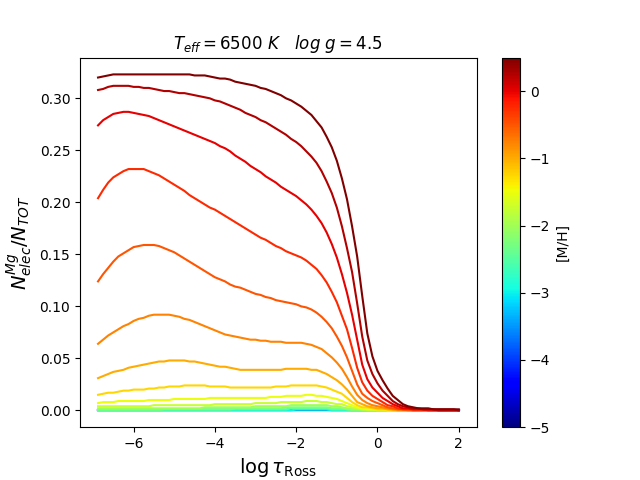}
\includegraphics[scale=0.55]{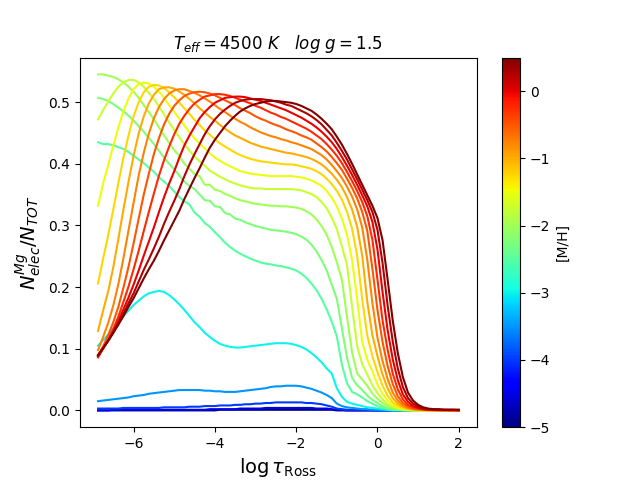}
\caption{Run of fraction of electrons provided by H (upper panels), 
and Mg (lower panels) as a function of \tros\ for the model atmospheres 
of a dwarf (\teff=6500 K and \gr=4.5, left panels) and a giant (\teff=4500 K and \gr=1.5, right panels) star, 
colour-coded according to the metallicity [M/H]. We adopt \alfa=+0.0 dex for all the models. }
\label{mod_elec}
\end{figure*}

For [M/H]$\lesssim$--3.0 dex, the gas pressure and the electron number density are very similar 
despite different metallicities in most of the model. In the outermost layers the metal-poor models 
have more steep gas pressures and higher electron number density.
This behaviour reflects the different contributor to free electrons provided by hydrogen and metals 
in the outermost layers of metal-poor atmospheres. As shown in Fig.~\ref{mod_elec}, 
showing the fraction of electrons provided by H and Mg as a function of \tros , 
in the external regions and [M/H]$<$--3.0 dex hydrogen becomes the dominant source of free electrons, 
while the contributor of Mg (the main electron donor for FGK stars) decreases significantly.

\section{Impact of the adopted \alfa\ on the model atmospheres }
\label{alfa}

The approach commonly used when computing a synthetic spectrum is to vary the abundance of a given element
while still using an atmospheric model that was calculated with a different abundance for that species. 
This procedure is legitimate and does not introduce significant errors as long as the abundance variations 
do not have a major impact on the atmospheric opacity or the ionization structure. Therefore, even
substantial changes in the abundances of elements that contribute little to the overall opacity do not lead
to inaccuracies in the resulting synthetic spectrum. On the other hand, greater caution should be exercised
when varying the abundances of elements such as the $\alpha$ elements. 
Therefore, the use of model atmospheres including appropriate chemical mixtures in terms of \alfa\ 
is recommended.

\begin{figure*}[ht!]
\centering
\includegraphics[scale=0.55]{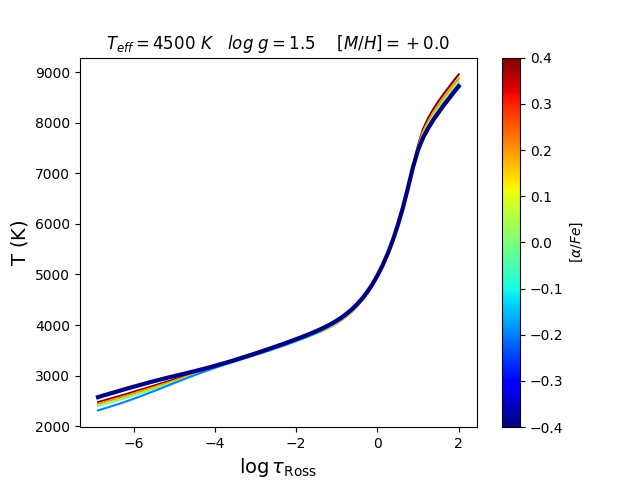}
\includegraphics[scale=0.55]{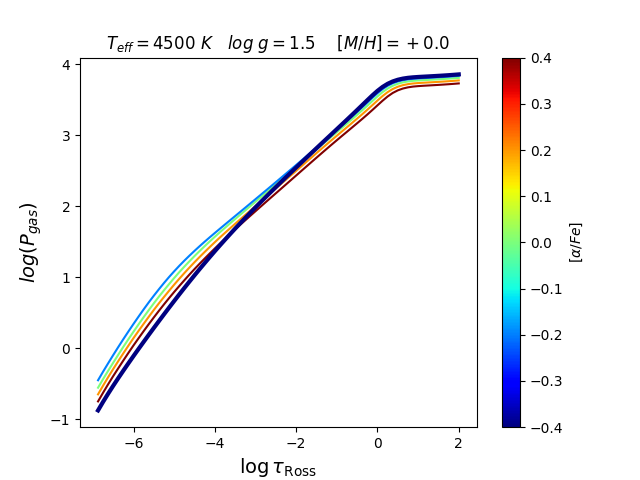}\\
\includegraphics[scale=0.55]{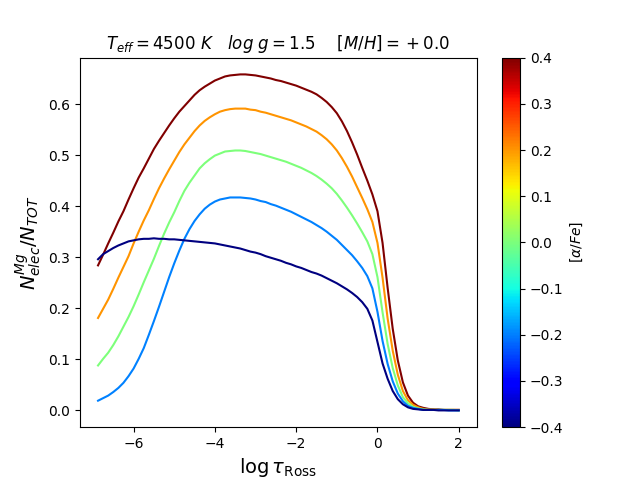}
\includegraphics[scale=0.55]{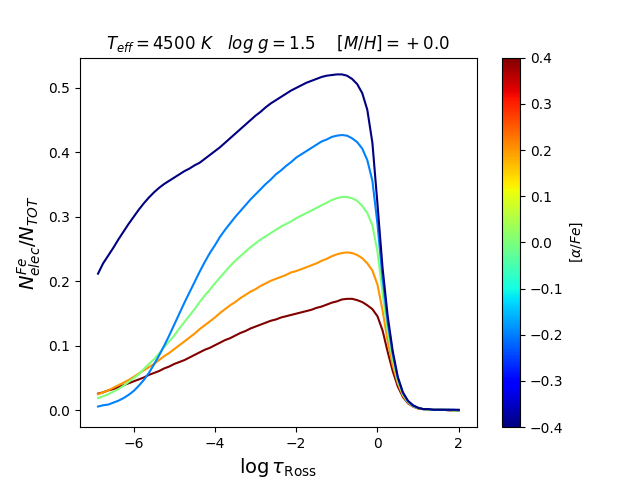}\\
\caption{Model atmospheres with \teff=4500 K, \gr=1.5, [M/H]=+0.0 dex and 
different \alfa . The panels show the behaviour of temperature, logarithm of the gas 
pressure, fractions of free electrons provided by Mg and Fe. 
The model with \alfa=--0.4 dex, showing a different behaviour with respect to the other ones, 
is plotted with a thicker curve.
}
\label{mod_alfa}
\end{figure*}

We checked the differences in the thermal and pressure structures of model atmospheres 
with different \alfa\ .
We recognized a peculiar pattern in cool models with [M/H]$\gtrsim$--2.0 dex and \alfa=--0.4 dex. 
As visible in Fig.~\ref{mod_alfa}, the decrease of \alfa\ at fixed [M/H] mimics a decrease of [M/H], 
with the features that we discussed above in Section~\ref{metallicity}. On the other hand, 
the model with \alfa=--0.4 dex is discrepant in the outermost layers, where it has a higher \teff\ 
and a lower lower $\rm P_{gas}$ than the model with \alfa=--0.2 dex (we should expect the opposite 
behaviour following Fig.~\ref{mod_teff}). This run could be explained by the peculiar chemical 
mixture of this model, where the extremely low \alfa\ changes the budget of the free electrons. 
In fact, we found that the free electrons in the outermost layers of this model arise mainly from 
Fe and Mg, both negligible in the external regions of the other models, where the majority 
of the free electrons are provided by Na and Al.

\begin{figure}[ht!]
\centering
\includegraphics[width=\hsize]{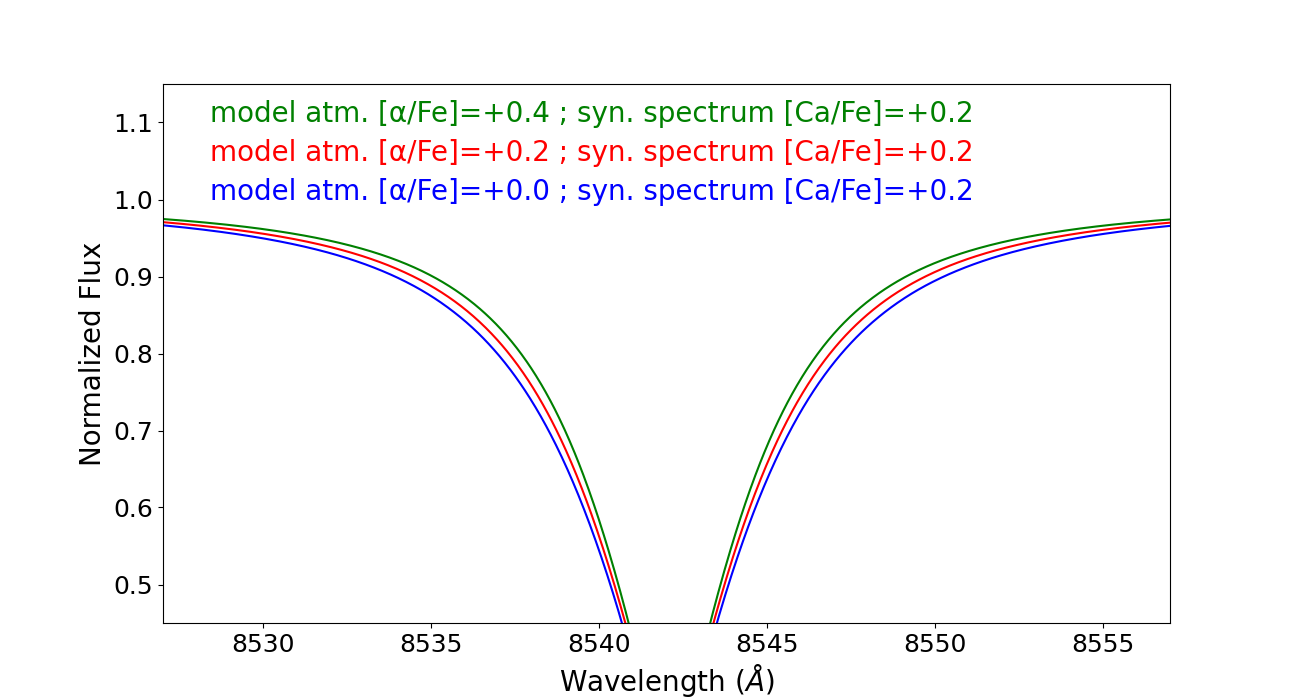}
\caption{Synthetic spectra for the second line of the Ca~II triplet calculated with 
model atmospheres with
\teff=~4500 K, \gr=~1.5, [M/H]=+0.0 dex and three different values of \alfa. 
Despite the latter value, all the synthetic spectra were computed assuming [Ca/Fe]=+0.2 dex.
}
\label{cat}
\end{figure}

A finer sampling of \alfa\ in the adopted model atmospheres has an impact 
also on the spectral synthesis, in particular on the strength of some specific features 
sensitive to the adopted chemical mixture of the model atmosphere.
An example of possible issues arising from use of model atmospheres with different \alfa\ 
is visible in the analysis of the Ca~II triplet lines. 
Fig.~\ref{cat} shows the comparison between synthetic spectra of the second Ca~II triplet line 
calculated with the code SYNTHE and using our new grid of model atmospheres. 
In particular, the three synthetic spectra were
calculated with the same Ca abundance ([Ca/Fe]=+0.2 dex) but assuming three different 
model atmospheres in terms of \alfa\, namely +0.0, +0.2, +0.4 dex. 
This means that in the spectral synthesis calculation we varied the Ca abundance 
of +0.2, 0.0 and --0.2 dex with respect to the original abundance of the models, respectively, 
in order to obtain the same [Ca/Fe].
Despite the same Ca abundance of the three synthetic spectra, the wings of the Ca~II line 
are different from each other. The broadening of these lines is dominated by van der Waals broadening 
and extremely sensitive to the gas pressure. The increase of [$\alpha$/Fe] in the model atmosphere 
(despite the Ca abundance variation adopted in the spectral synthesis) leads to a decrease of the 
gas pressure (similar to what happens when [M/H] increase) reducing the strength of the wings.
On the other hand, linear/saturated Ca lines are not affected by 
the assumption of the \alfa\ of the model atmosphere.

\section{Theoretical magnitudes and colours}

For each flux of the new grid described above we calculated theoretical magnitudes and colours 
in different photometric systems and bolometric corrections in Gaia DR3 G-band, BC(G). 
The procedure to calculate theoretical colours is described in details in Section~\ref{colours}. 
The photometric systems that we consider are the UBVRI \citep{bessel12}, the 2MASS JHK \citep{cohen03}, 
the Hipparcos-Tycho \citep{bessel12}, the SDSS {\sl ugriz} \citep{fukugita96}, the Euclid $\rm I_{E}Y_{E}J_{E}H_{E}$ 
\citep{euclid22}, the GALEX NUV and FUV \citep{morrissey07} and the Gaia DR3 photometric system \citep{gaia2023}. 

\begin{figure*}[ht!]
\centering
\includegraphics[scale=0.45]{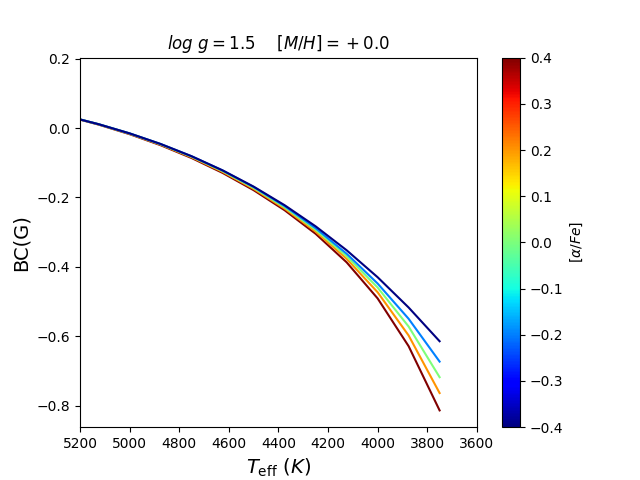}
\includegraphics[scale=0.45]{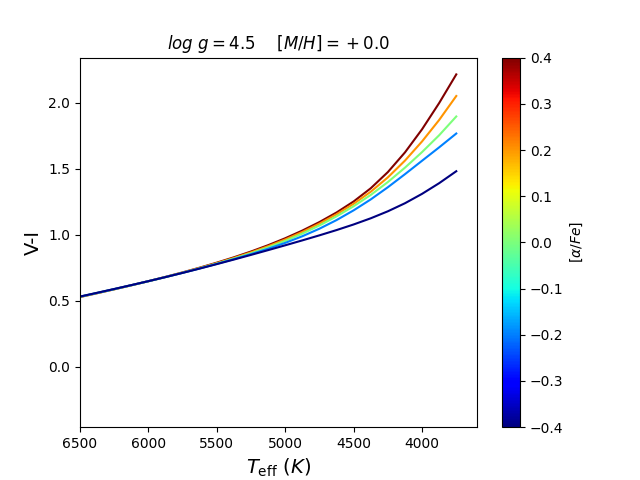}\\
\includegraphics[scale=0.45]{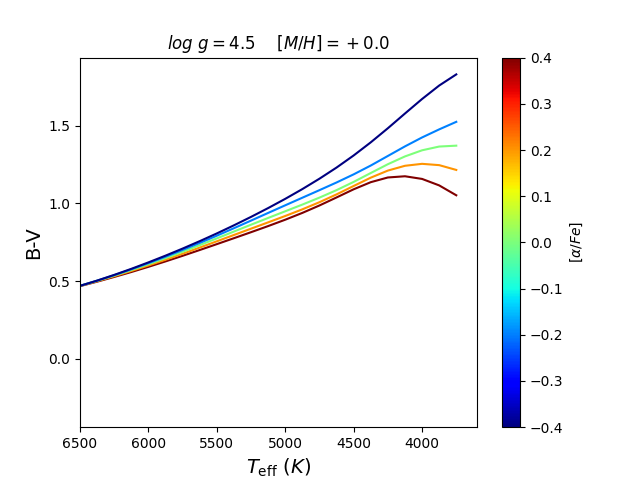}
\includegraphics[scale=0.45]{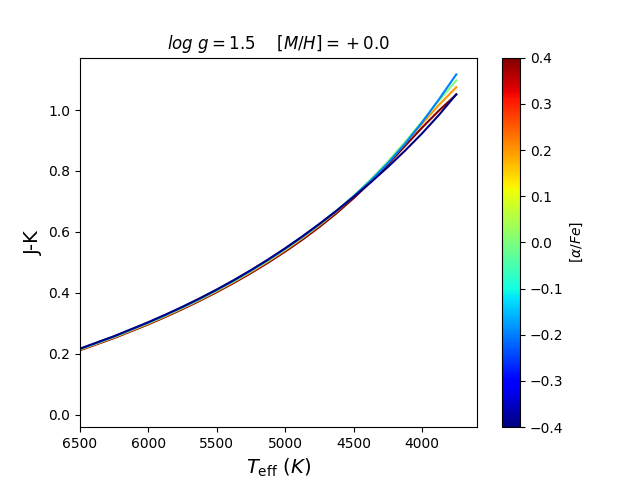}
\caption{Upper-left panel: run of the G-band bolometric correction as a function of \teff\ 
for model atmospheres with \gr=1.5, [M/H]=+0.0 dex and different \alfa . 
Upper-right panel: run of (V-I) as a function of \teff\ (\gr=4.5, [M/H]=+0.0 dex).
Lower-left panel: run of (B-V) as a function of \teff\ (\gr=4.5, [M/H]=+0.0 dex).
Lower-right panel: run of (J-K) as a function of \teff\ (\gr=1.5, [M/H]=+0.0 dex). 
}
\label{bcg}
\end{figure*}

The behaviour of bolometric corrections and colours as a function of the parameters, 
i.e. \teff, \gr, [M/H], has been extensively discussed in other works 
\citep[see e.g.][]{bessel98,bonifacio18,casagrande18}. Here we focus on their sensitivity 
on the adopted \alfa , the main novelty of this new dataset. 
The G-band bolometric correction BC(G) is basically unaffected by the adopted \alfa , but for 
metal-rich ([M/H]$\ge$+0.0 dex) giant models, for which BC(G) decreases by increasing \alfa . 
As example, Fig.~\ref{bcg} shows the run of BC(G) as a function of \teff\ for [M/H]=+0.0 dex and different values of \alfa . 
The sensitivity of BC(G) with \alfa\ can be appreciated 
at lower \teff .
At lower metallicities, the difference among the models, even at low \teff, decreases and disappears 
entirely. Below [M/H]$\sim$-–1.0 dex, BC(G) is completely independent of \alfa\ value.

\begin{figure*}[ht!]
\centering
\includegraphics[scale=0.55]{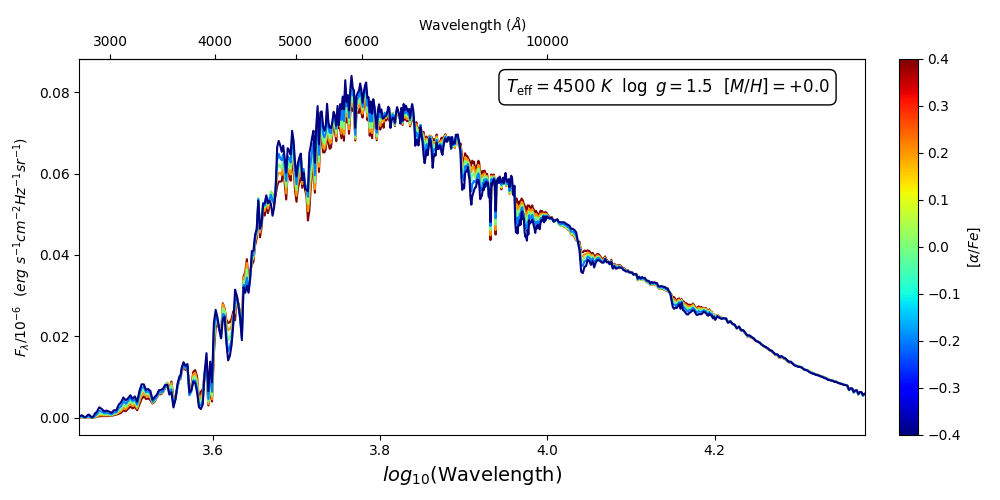}
\includegraphics[scale=0.55]{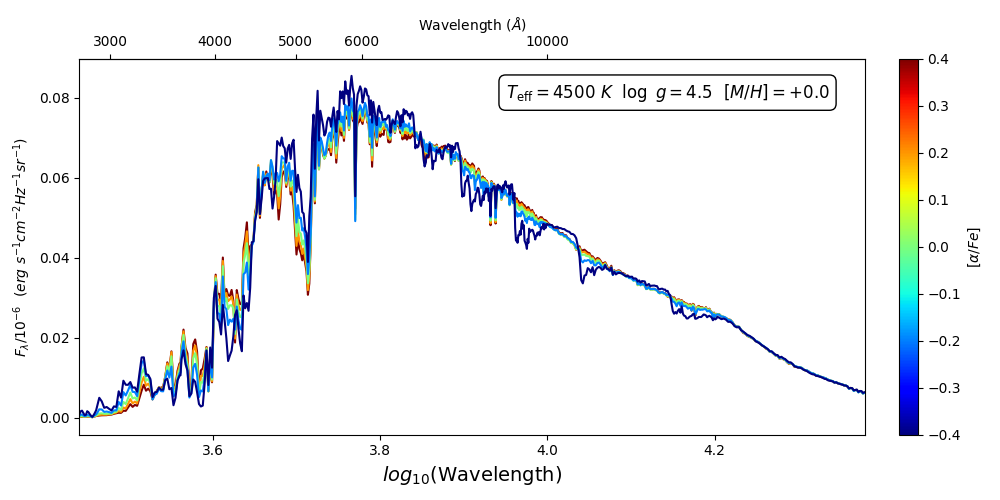}
\includegraphics[scale=0.55]{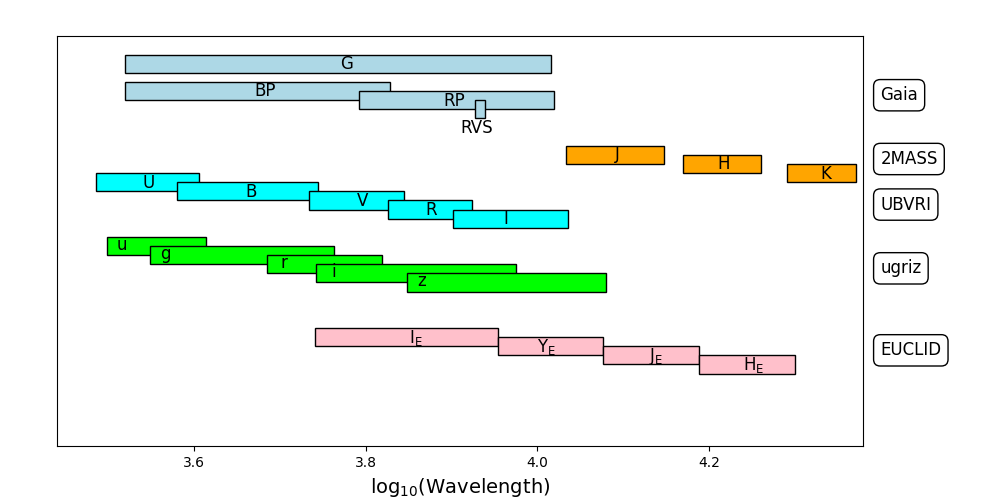}\\
\caption{Examples of ATLAS9 emergent fluxes calculated for a giant 
and dwarf star (upper and middle panel, respectively), calculated 
with different values of \alfa . 
The lower panel shows the log-wavelength range of the photometric filters 
discussed here.}
\label{fluxalfa}
\end{figure*}

Concerning the colours, the main effects of \alfa\ appear in general at low \teff\ and 
high [M/H]. 
The behaviour of colours with \alfa\, at fixed \teff\ and [M/H], is not univocal, as it depends on two opposing effects 
that affect the emerging spectrum when \alfa\ varies. On one hand, a decrease in \alfa\ naturally leads 
to a reduction in the strength of features associated with $\alpha$-elements, resulting in an increase in flux 
within the adopted filter profile, especially in the visual region between 
$\sim$4500 and $\sim$6500 \AA\ and at wavelengths shorter than $\sim$3200 \AA . On the other hand, the drop in oxygen alters the molecular equilibrium 
of the CNO cycle, progressively enhancing the strength of CN molecular bands \citep[see e.g.][]{ryde09}, at $\sim$3800 \AA\ and beyond 6500 \AA , and of the G-band CH band at $\sim$4300 \AA . This effect is significantly strong for models with \alfa=--0.4 dex, therefore the colours calculated with this chemical mixture are in almost all the cases the most divergent with respect to the other ones.
Fig.~\ref{fluxalfa} shows an example of these effects, with two emergent fluxes calculated 
for a K giant (\teff=4500 K, \gr=~1.5) and a K dwarf (\teff=4500 K, \gr=~4.5) star, 
both with [M/H]=+0.0 dex. In particular, the strong CN and CH molecular features are clearly visible 
in the emergent fluxes with \alfa=--0.4 dex.
Depending on the filters involved, these two effects can lead 
to either an increase or a decrease in the flux ratio as \alfa\ decreases.
In particular, we grouped the colours in three classes 
according to their sensitivity on the adopted \alfa\ :
\begin{itemize}
\item Colours directly proportional to \alfa : 
most of the colours become bluer by reducing \alfa\ , among them 
(r-i), (i-z), all the Gaia colours, (V-I) and ($\rm I_{E}$-$\rm Y_{E}$) both for giant and dwarf models, 
and (V-K) for dwarf stars. 
The run of (V-I) with \teff\ in dwarf models is shown in lower-right panel of Fig.~\ref{bcg} 
as example of this class of colours.
\item Colours inversely proportional to \alfa : 
other colours show an opposite behaviour, becoming redder by reducing \alfa , 
as (BT-VT), ($\rm Y_{E}$-$\rm J_{E}$) and, for dwarf stars, (U-B), (B-V), (u-g).
The run of (B-V) with \teff\ in dwarf models is shown in lower-right panel of Fig.~\ref{bcg}.
In these colours, the flux of the second filter is significantly reduced 
because the filter profile is dominated by the Mg b triplet (see Fig.~\ref{fluxalfa}).
\item Colours (almost or totally) insensitive to \alfa\ : 
all the other colours have negligible or lacking dependence on the adopted \alfa\ . 
Almost all the Euclid colours have little or negligible variations with \alfa\ 
but for the models with \alfa\=--0.4 dex where the colours are significantly 
discrepant with respect to other ones.
Lower-right panel of Fig.~\ref{bcg} shows as example the run of (J-K) with \teff .
\end{itemize}

\section{Summary}
We present a new database of ATLAS9 ODFs, model atmospheres, emergent fluxes, G-band bolometric corrections and 
theoretical magnitudes and colours. The latter have been calculated in different photometric filters, namely 
UBVRI, 2MASS, SDSS, Hypparcos-Tycho, Euclid, Galex and Gaia DR3. All the products are available in the dedicated website.
The new grid of ODFs includes a finer sampling in [M/H] (from --5.0 to +0.5 dex) and \alfa\ (from --0.4 to +0.4 dex). 
We discussed the impact of [M/H] and \alfa\ on model atmospheres (in terms of temperature, pressure and electron 
number density), emergent fluxes and theoretical colours. Among the features that we discussed, the most relevant 
are the following:
\begin{itemize}
\item 
models with [M/H]$\gtrsim$--2.5/--2.0 dex become 
increasingly distinct from each other at a fixed optical depth, both in terms 
of thermal and pressure structures, while more metal-poor models are often 
indistinguishable or show only small differences. This highlights the need 
of a finer sampling in [M/H] for model atmospheres with [M/H]$\gtrsim$--2.5/--2.0 dex;
\item the thermal structure of the model is sensitive to [M/H] 
in particular in deeper (\tros$\gtrsim$0.5) and outer (\tros$\lesssim$-3) layers. 
Also the gas pressure and the electron number density are significantly affected by [M/H] 
but in different and opposite ways;
\item 
the decrease of \alfa\ at fixed [M/H] impacts on the thermal and pressure structures of model atmosphere 
similar to the decrease of [M/H]. However, the models with \alfa=--0.4 dex is discrepant in the outermost layers, 
where it has a higher \teff\ 
and a lower lower $\rm P_{gas}$ than the model with \alfa=--0.2 dex. 
This is due to the large contribution of Fe and Mg to the budget of the free electrons 
in outermost layers;
\item 
The \alfa\ in the adopted model atmosphere can affect the spectral synthesis of some features 
dominated by the van der Waals broadening (like the wings of the Ca II triplet lines). 
Generally, it is recommended to adopt for the spectral synthesis a model atmosphere 
with an appropriate chemical composition in terms of \alfa ;
\item theoretical colours can have different sensitivity to \alfa , depending on 
the balance between two different effects affecting the involved filter profiles: 
the decrease in \alfa\ leads to weaker $\alpha$-element features (i.e. the Mg b triplet) 
and stronger CN and CH features. Again, the colours calculated with \alfa=--0.4 dex 
are often significantly discrepant with respect to the other ones.  

\end{itemize}

\begin{acknowledgements}
This work is dedicated to the memory of R. L. Kurucz, who passed away in March 2025, whose contribution to the study 
of stellar atmospheres was fundamental. 
We are grateful to R. Lallement for explaining to us the details of building
extinction maps and for reading a draft of our paper.
A.M. acknowledges support from the project "LEGO – Reconstructing the building blocks of the Galaxy 
by chemical tagging" (P.I. A. Mucciarelli) granted by the Italian MUR through contract PRIN 2022LLP8TK\_001. 

\end{acknowledgements}

%%%%%% TABELLE

\newpage
\makeatletter
\renewcommand{\@bibitem}[1]{\item[\hskip 0.5em\@biblabel{#1}]\hskip -0.5em}
\makeatother

{}

\begin{appendix}

\normalsize
\section{About the metallicity in ATLAS9 model atmospheres}
\label{metkur}
%%%%%%%%%%%%%%% STRUCTURE OF A MODEL
A model atmosphere is specified according to the effective temperature 
(\teff ,  that defines the total bolometric flux that traverses the photosphere 
through the Stefan-Boltzmann relation, $\rm F_{bol}=\sigma T_{eff}^{4})$, the surface gravity 
(\gr , that is related to the gas pressure through the hydrostatic equilibrium, 
$\frac{dP}{d\tau_{\nu}}=\frac{g}{k_{\nu}}$)
and the chemical composition. 
Therefore, it is not the overall metallicity that defines the characteristics of the model, but rather the details 
of its chemical mixture, particularly the abundance of those elements that have a significant impact on the total opacity 
(for instance the $\alpha$-elements).

ATLAS9 models tabulate explicitly the fraction of H and He atoms with respect to the 
total number of atoms ($\rm \frac{N_{H}}{N_{TOT}}$ and $\rm \frac{N_{He}}{N_{TOT}}$), while 
the metallic abundances are obtained by 
scaling the solar chemical composition (taken as reference) to a scaling factor 
(corresponding to the metallicity [M/H]).

In the ATLAS9 models, the abundance of the i-th metal is provided as $$\rm A_{kurucz}(Z_{i})=\rm log_{10}(\frac{N_{Z_{i}}}{N_{TOT}})$$
This formalism is slightly different to the traditional scale used to provide the absolute abundance 
of metals, namely $$\rm A_{trad}(Z_{i})=\rm log_{10}(\frac{N_{Z_{i}}}{N_{H}})+12$$ and we can easily translate 
the Kurucz scale into the traditional one as $$\rm A_{trad}(Z_{i})= A_{kurucz}(Z_{i})+12-\rm log_{10}(\frac{N_{H}}{N_{TOT}}).$$

\normalsize
\section{The zero-metallicity ODFs}
\label{zero}
Additionally, we compute a set of {\sl zero-metallicity} ODFs calculated without including 
the contribution of the metals but only hydrogen and helium.
This set is useful to calculate model atmospheres and fluxes for ideal Population III 
stars and provide boundaries for the photometric colours reliable for the search for 
very metal-poor stars. We compared some model atmospheres and corresponding fluxes for 
giant and dwarf Population III stars, finding that these models are indistinguishable 
from the most metal-poor ones of our grid. This exercise suggests that the spectral characterization 
of these rare, very metal-poor stars does not need additional, specific models.

%\newpage

\section{Computing theoretical magnitudes and colours}
\label{colours}

This topic has been often reviewed, and we refer the reader 
to \citet{bessel90,bessel98,castelli99,girardi04,bessel12,casagrande14,bonifacio17,bonifacio18} 
for definitions and discussion.
In this appendix we  detail how we computed magnitudes and colours presented in this paper.

All our codes have been derived from R. L. Kurucz's code
{\tt cousins}\footnote{\url{http://kurucz.harvard.edu/programs/colors/cousins.forcd}}, with one main difference,
while the original code performed energy integration 
\citep[eq. 3 of ][]{bonifacio17} we computed all
colours assuming photon counting.
Energy integration is
appropriate for $UBVRI$ photometry  
obtained with photomultiplier tubes operated in 
energy integration mode. In the last thirty years
however all photometry in any system has been 
obtained using photon counting detectors, including
photomultipliers operated in photon counting mode
\citep[e.g.][]{bonifacio00}.
The magnitude is computed as:

\begin{equation}
    m-m_0 = -2.5\log\left( 
    \frac{\int \lambda f(\lambda)R(\lambda)d\lambda}
    {\int\lambda R(\lambda) d\lambda}\right)
    \label{eq:mag}
\end{equation}

where $f(\lambda)$ is the flux from the star
and $R(\lambda)$ is the instrument response
function of the  photometric band, this 
must include not only the filter response, but
also the detector quantum efficiency and the telescope
throughput. For some systems like SDSS \citep{fukugita96},
also the transmission of the Earth's atmosphere is included.

From equation \ref{eq:mag} it appears that two
integrals have to be evaluated, however while for 
Vegamags, $m_0$ is a constant and the two integrals
appearing explicitly in equation\,\ref{eq:mag}
have to be evaluated, the situation is slightly
different for AB mags, like SDSS.

In this case $m_0$ is the magnitude of an object that has
a constant flux $F_\nu = 3631\times 10^{-23} {\rm ergs\,s^{-1}\,cm^{-2}\,Hz^{-1}} $.
Since the bandpasses of the filters are provided as a function of wavelength, it is convenient to carry out the integrations in wavelength, rather than frequency. In this case $ d\nu = \frac{c}{\lambda^2}d\lambda$ , where $c$ is the speed of light.

The AB magnitudes are defined as integrals over frequency, we use the above relation to transform
the integral over frequency into integral over wavelengths.
We follow what suggested in the Gaia DR3 
documentation\footnote{\url{https://gea.esac.esa.int/archive/documentation/GDR3/Data_processing/chap_cu5pho/cu5pho_sec_photProc/cu5pho_ssec_photCal.html}} equations 5.42 and 5.44 and $m_0=56.10$.

\begin{equation}
    m = -2.5\log \left(  \frac{\int \lambda f(\lambda)R(\lambda)d\lambda}
    {\int\frac{c}{\lambda} R(\lambda) d\lambda}\right) -56.10
    \label{mAB}
\end{equation}

The two integrals that appear in the numerator and denominator of
the right-hand of  equation \ref{mAB}  must be evaluated.
One warning: for SDDS we use magnitudes and not ``luptitudes'
as provided in the SDSS catalogue \citep{lupton1999}, the reasons for this choice are well explained in \citet{girardi04}.  
Following the original approach of R. L. Kurucz the integrals
are approximated by a sum of rectangles, where both
theoretical flux and instrumental response function have been
rebinned at a step of, typically, 0.1\,nm.
The instrument response function is rebinned using the polynomial
interpolation subroutine {\tt PINTER} from Kurucz's original code, 
and the theoretical flux is linearly interpolated
using the ATLAS subroutine {\tt LINTER}. For some filters, like e.g. SDSS $z$,
PINTER provides some negative values at the very edge of the response
function, in this case we simply set the response function to zero.
The SDSS $z$ filter is rebinned at steps of 0.5\,nm that is sufficient
to recover the correct flux.

Another difference with respect to Kurucz's original code is that
we adopt the approach of \citet{casagrande14}: we 
compute the magnitude for every model as if it corresponds to a star
of  one solar radius at a distance of 10 pc.
In practice we multiply the theoretical
fluxes by a dilution factor $(R_\odot/10)^2= 5.083267\times10^{-18}$.

ATLAS fluxes are provided as $H_\nu$ in units
$\rm erg\, s^{-1}\, cm^{-2}\, Hz^{-1}\,sr^{-1}$, we transform this to $F_\lambda$
so we need to insert a factor $4\pi$
and since we express wavelengths in nm, we transform the units
to $\rm W m^{-2} nm^{-1}$
in practice for each wavelength 
WAVE(NU) in the theoretical flux HNU(NU)
we use this fragment of {\tt FORTRAN} code to perform
the transformation
\begin{verbatim}
FREQ=2.99792458E17/WAVE(NU)
HLAM(NU)=1.e-3*HNU(NU)*FREQ/WAVE(NU) * DIL *(4*PI)
\end{verbatim}

where DIL is the above-defined dilution factor,
PI is $\pi$ and HLAM is $F_\lambda$.

The theoretical fluxes are also useful to define the 
extinction coefficients in the various bands.
The relationship between the intrinsic magnitude of
a star and the magnitude observed by us is:
\begin{equation}
    m_i = m+5\log \left(\frac{R_\ast}{d}\right)-A
    \label{eq:extinction}
\end{equation}

where $m_i$ is the intrinsic magnitude, $m$ is the 
observed magnitude, $R_\ast$
is the stellar radius, $d$ is the distance of the star
and $A$ is the total extinction. If the magnitudes 
are monochromatic at wavelength $\lambda$ $A$ is the extinction at that wavelength.
If the magnitudes are defined over a band, $A$ is the integral of the monochromatic
extinction over the band.
Equation\,\ref{eq:extinction} can also be interpreted as {\em definition}
of total extinction. It is clear by combining equation\,\ref{eq:mag}
and equation\,\ref{eq:extinction}
that $A$ is a function of the flux distribution of the star.
This means that the light of stars of different \teff\ , \gr\ and metallicity
going through the same interstellar medium will have a different $A$.
Since, in general,  the angular diameter of stars is not known 
it is convenient to express extinctions as ratios with respect
to extinction at a given wavelength or in a given band.
The two, by far, most common choices for reference extinctions are $A_V$, the extinction
in Johnson's $V$ band and $A0$, the monochromatic extinction at 550\,nm.
There are many 2D \citep[e.g.][]{schlafly11} and 3D Galactic maps
\citep[e.g][]{schlafly14,vergely22} that provide extinction as a function
of position on the sky (2D) or position on the sky and distance (3D).
The map of \citet{vergely22} is derived from the  reddening estimates
of a large number of stars, obtained with different methods, both photometric
and spectroscopic. In a first step
the derived $A_0$ for any of the stars is indeed a function of the stellar parameters. 
In a second set all the the different estimates are inter-calibrated and combined
with the basic assumption that for the stars in a small space volume.
The third, and quite complex, step 
in making an extinction map
is the 
geometric computation to convert all sun-star integrated quantities to 
local values, this process is called inversion \citep{lallement14,vergely22}.
The 
properties of the dust are the same (or equivalently that this volume is homogeneous)
and from this process one determines a value of $A_0$ that is only a function
of the spatial coordinates and not of the stellar type.
All reddening maps need to make an assumption
of this type, that is some homogeneity of the 
dust in a small volume of space. 
Once the inversion is made and maps are computed, they provide an 
estimate of the extinction (e.g. $A_0$) in each point in three dimensional space 
that is fully independent of target star properties, provided that the 
individual determinations have used correct assumptions on stellar 
parameters/spectra. However, it is only the value of the extinction at 
the exact wavelength used for the mapping which is a property of the 
interstellar dust.  If , for a given observed star, one wants to use a 
map value, e.g. if one wants to use $A_0$, but would like, also, to have 
estimates of the extinction at other wavelengths or in a photometruc band, 
one should be aware that only the extinction at other wavelengths depends on 
the stellar properties. 

If one wants to correct for reddening the photometry of a star,
one should know the parameters of the star. Either one uses a method
independent of photometry to determine the stellar parameters, or one
uses an iterative procedure such as described, for example, 
in \citet{bonifacio19} or \citet{lombardo21}.
Since in reality the extinction coefficients do not 
have a strong dependence on the stellar parameters, the strongest
one being on effective temperature \citep[see e.g.][]{danielski18},
it is often acceptable, to use a single extinction coefficient
for all the stars. This is usually the approach adopted in 
dereddening the photometry of star clusters and dwarf galaxies.

In this paper for any synthetic band, $X$ provided we also
provide the ratio $AX/A_0$, if \teff, \gr , metallicity and
[$\alpha$/Fe] of the star are known, even approximately,
it is straightforward to interpolate in our tables to determine
the appropriate extinction, that can be be combined using a
value of $A_0$ obtained from maps or otherwise.
Since several maps are provided in terms of $A_V$ or $E(B-V)$, we also
provide the ratio $A_V/A_0$ that can be used to convert between the two extinctions.

In order to compute the extinction coefficients we had to assume
a Galactic extinction law and we assumed that of
\citet{fitzpatrick2019} corresponding to $R(55)=3.02$\footnote{$R(55)=\frac{A(55)}{E(44-55}$, 
the value 3.02 corresponds to $R(V)= \frac{A_V}{E(B-V}$= 3.1 .}.
The extinction cast into flux units,
using 
\begin{equation}
    b = 10^{-\frac{k+3.02}{2.5}}
    \label{eq:b}
\end{equation}

for any wavenumber, then the wavenumbers were transformed
in wavelengths and each vector sorted.
The value of $b_0$ at 550\,nm was obtained with a Hermite
spline interpolation \citep[subroutine INTEP,][]{hill82}
and we then computed $A_0 = -2.5\log(b_0)$
For any filter the the extinction law was rebinned with PINTER on the
same mesh as the filter response function.
For each filter we computed
\begin{equation}
  A=  -2.5\log\left( 
    \frac{\int \lambda f(\lambda)R(\lambda)b(\lambda)d\lambda}
    {\int\lambda f(\lambda)R(\lambda) d\lambda}
    \right)
\end{equation}

what is provided in the tables is $A/A_0$.

One final comment, all our synthetic photometry is provided
with six significant figures (five decimal places), not because
we believe that this is its accuracy, but because this helps
to minimise interpolation errors when using these tables.

\end{appendix}

\end{document}